\documentclass[preprint,aps]{revtex4}
\usepackage{amsmath}
\usepackage{graphicx}

\def\be{\begin{equation}}
\def\ee{\end{equation}}
\def \bea#1\eea {\begin{eqnarray}#1\end{eqnarray}}

\begin{document}


\title{A Family of Tunable Spherically-Symmetric Potentials that Span the
Range from Hard Spheres to Water-like Behavior
}

\author{Zhenyu Yan$^1$, Sergey V. Buldyrev$^{2,1}$, Nicolas
Giovambattista$^{3}$, Pablo G. Debenedetti$^{3}$ and H. Eugene Stanley$^1$}

\affiliation{
$^1$Center for Polymer Studies and Department of Physics, Boston
University, Boston MA 02215, USA\\
$^2$Department of Physics, Yeshiva University, 500 West 185th Street,
  New York, NY 10033 USA\\
$^3$Department of Chemical Engineering, Princeton University,
Princeton, New Jersey 08544-5263 USA
}

\date{\today}

\begin{abstract}

We investigate the equation of state, diffusion coefficient, and structural order
of a family of spherically-symmetric potentials consisting of a hard core and a
linear repulsive ramp. This generic potential has two characteristic length scales:
the hard and soft core diameters. The family of potentials is generated by varying
their ratio, $\lambda$. We find negative thermal expansion (thermodynamic anomaly) and an
increase of the diffusion coefficient upon isothermal compression (dynamic anomaly)
for $0\leq\lambda<6/7$. As in water, the regions where these anomalies occur are nested domes
in the ($T, \rho$) or ($T, P$) planes, with the thermodynamic anomaly dome contained
entirely within the dynamic anomaly dome. We calculate translational and
orientational order parameters ($t$ and $Q_6$), and project equilibrium state points onto
the ($t, Q_6$) plane, or order map. The order map evolves from water-like behavior to
hard-sphere-like behavior upon varying $\lambda$ between 4/7 and 6/7. Thus, we traverse the range of
liquid behavior encompassed by hard spheres ($\lambda=1$) and water-like 
($\lambda\sim4/7$) with a
family of tunable spherically-symmetric potentials by simply varying the ratio of
hard to soft-core diameters. Although dynamic and thermodynamic anomalies occur
almost across the entire range $0\leq\lambda\leq1$, water-like structural anomalies (i.e., 
decrease in both $t$ and $Q_6$ upon compression and strictly correlated $t$ and $Q_6$ in the 
anomalous region) occur only around $\lambda=4/7$. Water-like anomalies in structure, dynamics 
and thermodynamics arise solely due to the existence of two length scales,
orientation-dependent interactions being absent by design.

\end{abstract}


\maketitle

\section{Introduction}

Most liquids become denser when cooled and more viscous when compressed.
In contrast, water becomes less dense when cooled (density, or
thermodynamic, anomaly) and its diffusivity increases upon pressurization
(diffusion, or dynamic, anomaly). These anomalies, which disappear at high
enough temperature and pressure, are not unique to water. Other liquids
with local tetrahedral order (e.g., silica and silicon) also exhibit
thermodynamic and dynamic anomalies~\cite{angellPCCP}. A possible explanation of these
anomalies is the tendency of these substances to form local open
structures not present in simple liquids. However, establishing a precise
and quantitative link between the microscopic structure and the dynamic
and thermodynamic anomalies of tetrahedral liquids has proved elusive
until recently.

Errington and Debenedetti~\cite{jeffrey01} (ED) studied the relation between
microscopic structure and the anomalies of liquid water by introducing
two simple metrics: a translational order parameter 
$t$~\cite{errington03}, quantifying the tendency of molecule pairs 
to adopt preferential separations, and an orientational order parameter 
$q$~\cite{chau98,jeffrey01}, quantifying the
extent to which a molecule and its four nearest neighbors adopt a
tetrahedral local structure, as in the case of hexagonal ice. A useful way
of investigating structural order in liquids is to map state points into
the $t-q$ plane. Such a representation was introduced by Torquato and
coworkers~\cite{torquato00}, who first applied it to sphere packings and 
referred to it as an order map. ED used the order map to investigate 
structural order in water~\cite{jeffrey01}. Because of the distinctive 
features discovered in that study,
in what follows we refer to water-like order maps as the ED order map.
Using molecular dynamics simulation of the SPC/E~\cite{spce} model, ED found
that the state points accessible to liquid water define a two-dimensional
region in the $t-q$ plane, meaning that in general $t$ and $q$ are independently
variable in liquid water (i.e., equilibrium state paths exist along which
one order metric varies while the other does not). They also found a
dome-shaped region in the ($T,\rho$) plane within which isothermal compression
leads to a decrease in $t$ and $q$. This {\it decrease} in order upon compression
constitutes a structural anomaly: simple liquids, in contrast, always
become more ordered upon compression. ED further found that dynamic and
thermodynamic anomalies define nested domes in the ($T, \rho$) plane: the
structural anomalies dome contains the dynamic anomalies dome, which in
turn contains the thermodynamic anomalies dome. This means that whenever
the thermal expansion coefficient is negative, the diffusivity must
necessarily increase upon isothermal compression. ED showed that all state
points exhibiting structural, dynamic or thermodynamic anomalies define a
line on the ($t, q$) plane, meaning that when water exhibits anomalous
behavior, its translational and orientational order metrics become
strictly coupled. This is clear evidence of the relationship between
structure and water anomalies. Shell {\it et al.} subsequently found
qualitatively similar behavior in molten silica's order map~\cite{shell02}. 
However, in the case of silica, it was found that state points corresponding to
anomalous behavior define a narrow stripe in the ($t, q$) plane instead of a
strict line. Furthermore, unlike in water, the region of dynamic anomalies was found to
contain that of structural anomalies.

For simple spherically-symmetric liquids, including hard 
spheres~\cite{torquato00,truskett00} and Lennard-Jones~\cite{errington03} 
the order map was found to be a positively-sloped
line in the ($t, q$) plane, indicating that translational and orientational
order are always strictly and positively correlated. In this case, of
course, the appropriate metric for orientational order does not measure
tetrahedrality; rather, the bond-orientational order parameter introduced
by Steinhardt {\it et al.}~\cite{steinhardt83} was used. 
An important result from these studies
is the fact that the order map for the Lennard-Jones system above its
critical density is identical to that of hard spheres. Furthermore, in
these simple systems that do not exhibit thermodynamic or dynamic
anomalies, compression always leads to an increase in the order metrics.

In 1970 Hemmer and Stell~\cite{Stell70} showed that in fluids interacting via
pairwise-additive, spherically-symmetric potentials consisting of a hard
core plus an attractive tail, softening of the repulsive core can produce
additional phase transitions. This pioneering study elicited a
considerable body of work on so-called core-softened potentials
\cite{Stell70,stillinger97,Jagla99,Jagla01,Sadr98,Scala01,franzese,predpreprint,pablo91}.
This generic term denotes
continuous potentials with inflections in the repulsive core~\cite{stillinger97},
discontinuous potentials with the core softened by shoulders or ramps
\cite{Stell70,Jagla99,Sadr98,Scala01,franzese,predpreprint}, 
or lattice models with nearest-neighbor attraction
and next-nearest neighbor repulsion~\cite{pablo91}. It is now well-established that
such potentials can generate water-like density and diffusion anomalies
\cite{Stell70,stillinger97,Jagla99,Jagla01,Sadr98,Scala01,franzese,predpreprint,pablo91}.
This important finding implies that strong orientational
interactions, such as those that exist in water and silica, are not a
necessary condition for a liquid to have thermodynamic and dynamic
anomalies.

The above discussion implies the existence of two well-defined classes of
liquids: simple and water-like. The former which interact via
spherically-symmetric non-softened potentials, do not exhibit
thermodynamic nor dynamic anomalies, and their order map is a line. In
water-like liquids, interactions are orientation-dependent; these liquids
exhibit dynamic and thermodynamic anomalies, and their order map is in
general two-dimensional but becomes linear (or quasi-linear) when the
liquid exhibits structural, dynamic or thermodynamic anomalies.
Intermediate between these well-defined extremes is the class of
core-softened liquids, which interact via spherically-symmetric potentials
but can also exhibit water-like thermodynamic and dynamic anomalies.

Two questions arise naturally from this emerging taxonomy of liquid
behavior. First, is structural order in core-softened fluids hard-sphere
or water-like ? Second, is it possible to seamlessly connect the range of
liquid behavior from hard spheres to water-like by a simple and common
potential, simply by changing a physical parameter ?

In a recent study~\cite{zyan} we addressed the first question. We showed that a
core-softened potential with two characteristic length scales not only can
gives rise to water-like diffusive and density anomalies, but also to an
ED water-like order map. This implies that orientational interactions are
not necessary in order for a liquid to have structural anomalies. In this
work we address the second question. Specifically, we use the ratio of
characteristic length scales as a control parameter to investigate the
evolution of dynamic, thermodynamic and structural anomalies. In this
manner we show that the family of tunable spherically-symmetric potentials
so generated evolves continuously between water-like and hard sphere
behavior. To our knowledge this is the first time that essential aspects
of the wide range of liquid behavior encompassed by hard spheres and
tetrahedrally-coordinated network-formers can be systematically traversed
by varying a single control parameter.

This paper is structured as follows. Sections II, III, and IV provide
details on the interaction potential, simulation method, and order
parameters, respectively.  Results are presented in Section V. Conclusions
and some suggestions for future work are provided in Section VI.

\section{Ramp potential}

\begin{figure} \includegraphics[width=12cm]{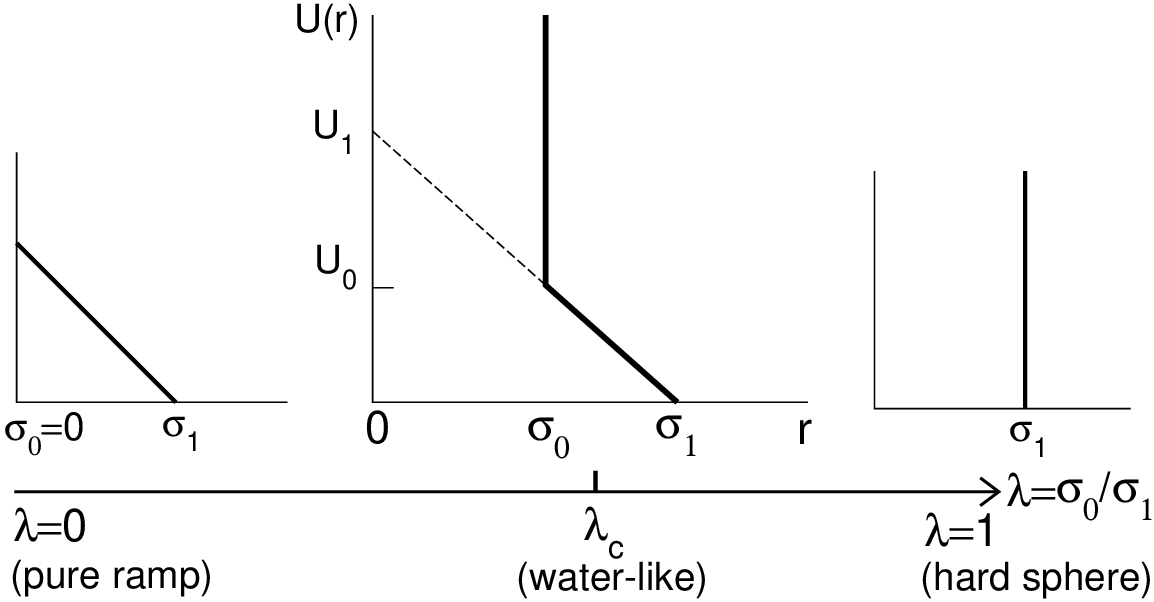} 
\caption{ 
The middle figure shows the ramp potential with two characteristic length scales. 
$\sigma_0$ corresponds to 
the hard core, $\sigma_1$ characterizes the onset of soft repulsion.
When $\lambda=0$ (left figure) we have a pure ramp 
potential (no hard core).
When $\lambda=1$ (right figure) we have a hard sphere potential. $\lambda_c\sim0.6$ is the 
ratio near which the system exhibits water-like structural, dynamic and thermodynamic
behavior. 
} 
\label{ramppotential} 
\end{figure}

We perform discrete MD simulations 
to study the equation of state, diffusion coefficient and structural order 
as measured by the ED order map, for a fluid whose particles interact via a 
pairwise-additive, spherically-symmetric potential that gives rise to 
both thermodynamic and dynamic water-like 
anomalies. The model was introduced by Jagla~\cite{Jagla99}; the
potential energy $U(r)$ between a pair of particles separated by a distance $r$ 
is given by (see Fig.~\ref{ramppotential}).
\begin{equation}
U(r)=\left\{
\begin{array}{ll}
\infty & r < \sigma_{0}\\
U_1(\sigma_1-r)/\sigma_1 & \sigma_0< r < \sigma_1\\
0 & r>\sigma_1
\end{array}\right.
\label{eq:ramp}
\end{equation}
The shorter distance $\sigma_0$ corresponds to the hard core,
and the longer distance $\sigma_1$ characterizes a softer repulsion
range that can be overcome at high pressure. Because of its shape, 
it is called the ramp potential. The constant slope of the 
ramp potential for $\sigma_0<r<\sigma_1$ keeps the force between 
particles $f$ constant, so the
product of separation and force $rf$ will decrease when the separation $r$
decreases. This satisfies the mathematical meaning of 
core-softening~\cite{pablo91} and under these conditions the thermodynamic
(density) anomaly can be 
qualitatively explained by invoking the virial theorem~\cite{pablo91}. 

Of interest is the ratio between the two characteristic length scales, 
$\sigma_0$ and $\sigma_1$, 
\begin{equation}
\lambda\equiv\sigma_0/\sigma_1,
\label{eq:ratio}
\end{equation}
which can vary
between 0 and 1. Ref.~\cite{zyan} investigated the one-scale ($\lambda=0$) 
and two-scale ($\lambda=4/7$) ramp potentials.
Here we investigate the full range $0<\lambda<1$, with $\lambda=0, 2/7, 4/7, 5/7, 6/7$.

\section{MD simulation}

We use discrete MD simulation; details are given in
Ref.~\cite{predpreprint}. We use the NVT ensemble for a system 
composed of $850$ particles with periodic boundary conditions, 
and we control the temperature with the Berendsen thermostat~\cite{allen}. 
However, we note that we use different units
than in Ref.~\cite{predpreprint}: the distance $r$, number density $\rho$, 
pressure $P$ and temperature $T$ are all normalized with respect to the 
soft core distance $\sigma_1$ and the potential $U_1$ at $r=0$. 
We also investigate systems with different number of particles ($N=1728$) 
and confirm that the results do not depend on the number of particles. 
The simulation ranges of temperature and density fully cover the region
where density, diffusion and structural anomalies occur.

\section{Translational and orientational order parameters}

\subsection{Translational order parameter}

The translational order parameter~\cite{jeffrey01,errington03,shell02} is defined as,
\begin{equation}
t\equiv\int_{0}^{s_c}{\lvert}g(s)-1{\rvert}ds.
\label{top}
\end{equation}
Here $s{\equiv}r\rho^{1/3}$ is the radial distance $r$ scaled by the
mean interparticle distance $\rho^{-1/3}$, $\rho$ is the number density,
$g(s)$ is the pair correlation function, and $s_c$ a
numerical cutoff. We choose $s_c$ so
that it corresponds to one-half the simulation box size, and we verify that 
our system size is always large enough so that $g(s)=1$ 
at half the box size. For a
completely uncorrelated system, $g(s)\equiv 1$, and thus $t=0$.
For systems with long-range order, the modulations in $g(s)$ persist over
large distances, causing $t$ to grow. Between these limits, $t$ will change
as a consequence of the dependence of $g(r)$ upon $T$ and $\rho$.

\subsection{Orientational order parameter}

An orientational order parameter based on spherical harmonic function was 
introduced by Steinhardt {\it et al.}~\cite{steinhardt83} and used in Refs.
\cite{errington03,torquato00,truskett00,huerta04}.
In this definition, all vectors connecting nearest neighbors 
(i.e., particle pairs whose separation is
less than the first minimum of the radial distribution function) are
considered. Each of these vectors, also called `bond', defines an
azimuthal and polar angle, and the corresponding spherical harmonic
function is evaluated. The orientational order parameter used in Refs.
\cite{errington03,torquato00,truskett00,huerta04}
involves the average of each spherical harmonic function over
all bonds.

The orientational order parameter used for water and silica in Refs.
\cite{jeffrey01,shell02} involves, first, the evaluation of the local tetrahedral order for
each particle with respect to its four nearest neighbors, and then, the
average of this quantity over all the molecules of the system. In the
definition of the orientational order parameter used in Refs.
\cite{errington03,torquato00,truskett00,huerta04}
there is no such concept of `local order' for an individual particle.
Moreover, the number of bonds associated to each particle is not fixed,
but instead it changes with temperature and pressure.
These two differences led, in Ref.~\cite{zyan}, to the introduction of a
slightly modified version of the original orientational order parameter
introduced by Steinhardt et al. in ref. [9]. The resulting order metric
is based on the idea of a `local order' for each particle, analogous to
Refs.~\cite{jeffrey01,shell02}. The ED maps obtained with the original (global) and modified
(local) definitions of orientational order are qualitatively similar.

In this work we use the same order parameter introduced in Ref.~\cite{zyan}. We
define twelve bonds connecting each particle with its twelve nearest
neighbors. Each bond is characterized by its azimuthal and polar angles
$(\theta,\varphi)$ and the corresponding spherical harmonic $Y_{{\ell m}}(\theta,\varphi)$ is
computed. The orientational order parameter associated with each
particle $i$ is
\begin{equation}
Q_{{\ell}i}\equiv\left[\frac{4\pi}{2\ell+1}\sum_{m=-\ell}^{m=\ell}\vert\overline{Y}_{\ell
m}\vert^2 \right]^\frac{1}{2}.
\label{q6}
\end{equation}
Here, $\overline{Y}_{\ell m}(\theta,\varphi)$
denotes the average of $Y_{\ell m}(\theta,\varphi)$
over the 12 bonds associated with particle $i$.
For $\ell=6$~\cite{errington03}, $Q_{{\ell}i}$ has a large
value for most crystals such as fcc, hcp and bcc~\cite{steinhardt83}.
The values of $Q_{6i}$
for each molecule in the system follow a Gaussian distribution. $Q_6$, the 
averaged value of $Q_{6i}$ over all particles $i$, 
is used to characterize the local order of the system.
This definition of order parameter is analogous to that used in water.
For water, the solid at low pressure is hexagonal ice where each
molecule has four neighbors. The orientational order parameter is
maximum in the ice configuration and decreases as the system becomes
less ice-like. For the ramp potential, the solid phase at low pressure
has a fcc structure where each particle has twelve nearest neighbors. $Q_6$
has a maximum value in the fcc lattice ($Q_{6}^{\rm fcc}=0.574$) and decreases
as the system becomes less correlated (for uncorrelated systems,
$Q_{6}=1/\sqrt{12}=0.289$).

\section{Results and Discussion}

\subsection{Structural anomalies and order map}

\begin{figure} \includegraphics[width=12cm]{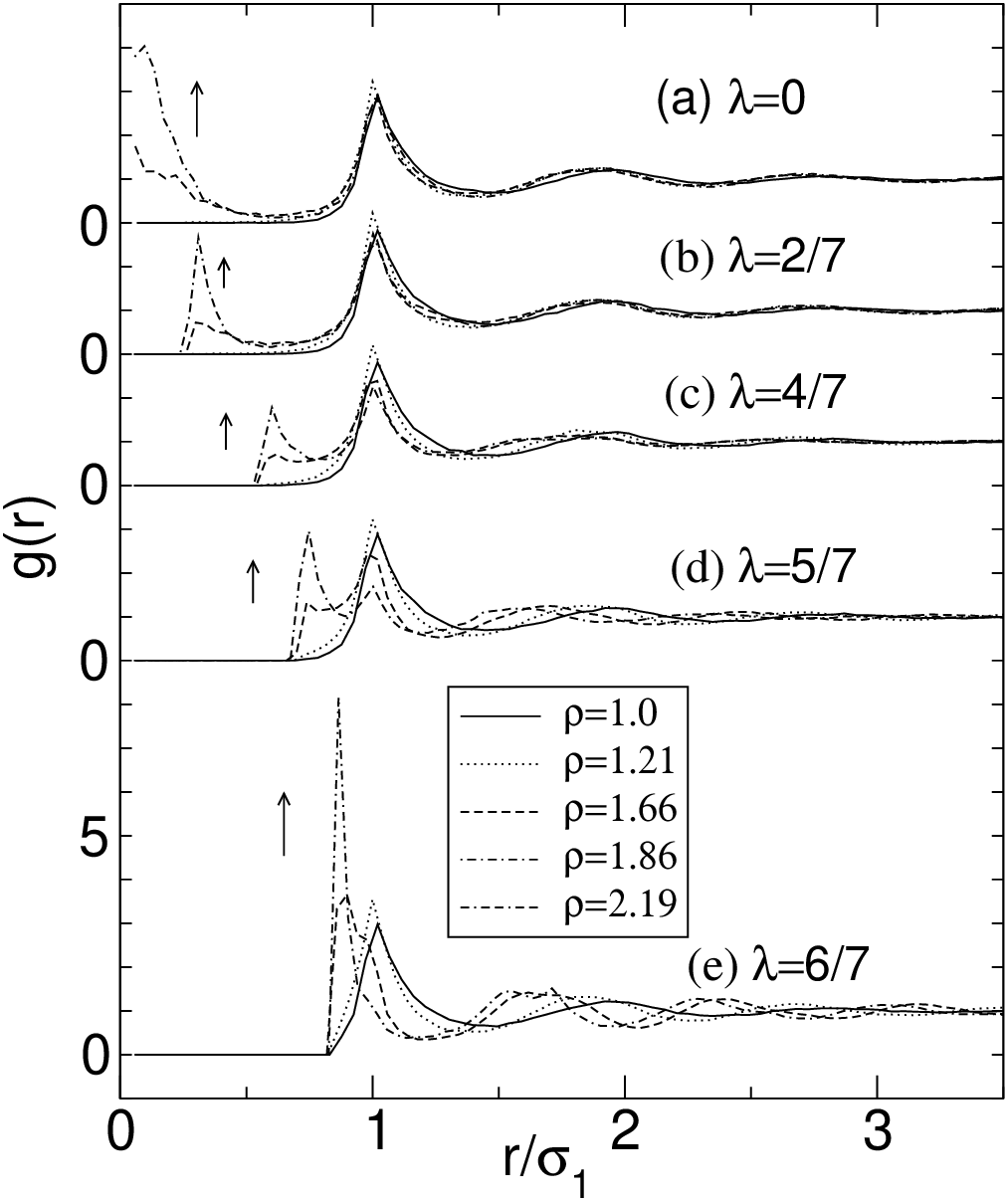} 
\caption{
Radial distribution function $g(r)$ at $T=0.04$ for different
$\lambda\equiv\sigma_0/\sigma_1$ values.
The arrows indicate the direction of increasing density.
The density values are 1.0, 1.21, 1.66, 2.19 for $\lambda=2/7,4/7,5/7$ 
and 1.0, 1.21, 1.66, 1.86 for $\lambda=0, 6/7$. 
The distance $r$ is normalized by $\sigma_1$, the soft core length.
The curves are shifted vertically for clarity.
} 
\label{rdfs} 
\end{figure}

Since the pair correlation function $g(r)$ is used to compute the translational
order parameter $t$ (Eq.~\ref{top}), we first discuss the effect of density on $g(r)$. 
Fig.~\ref{rdfs} shows the effects of compression on $g(r)$ at low
temperature, $T=0.04$, for the various values of $\lambda$ considered in this study. In
all cases, there is no inner peak at $r=\sigma_0$ for $\rho=1$ and 1.21, and only the outer
peak at $r=\sigma_1$ is present at these densities. The inner peak at $r=\sigma_0$, which is
broad and of modest height at $\rho=1.66$, becomes sharper and more pronounced upon
further compression. Interestingly, structural changes brought about by
compression become progressively longer-ranged as $\lambda$ increases. Thus, for $\lambda=0$ and
2/7, the major changes in $g(r)$ involve the development of structure at length
scales $\leq\sigma_1$ associated with the growth of the inner peak at $r=\sigma_0$. However, for
$\lambda=4/7, 5/7$ and 6/7, structural changes upon compression occur at distances larger
than $\sigma_1$. In particular, for $\lambda=6/7$, the effects of compression are clearly
discernible at $r=3\sigma_1$. 

\begin{figure}
\includegraphics[width=16cm, height=7.2cm]{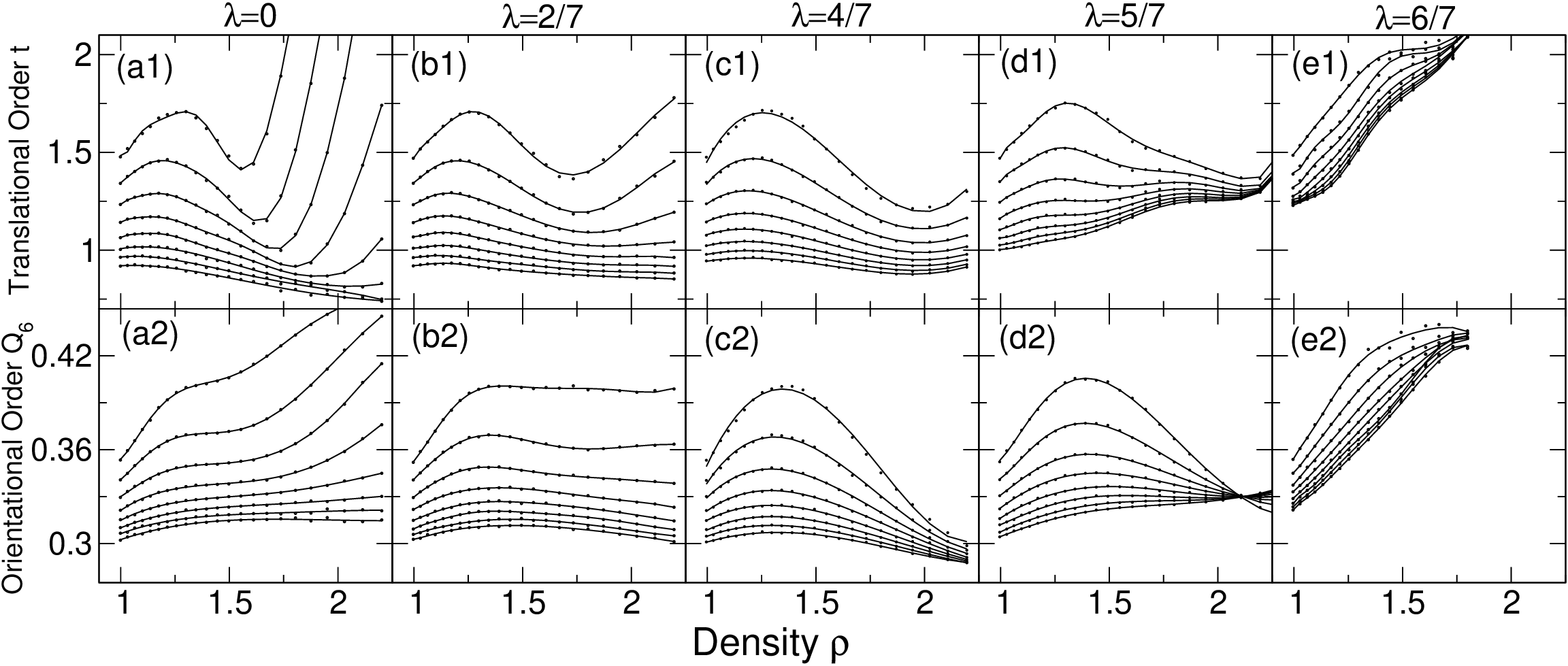}
\caption{
 The upper panels (a1)--(e1) show the density-dependence of the
 translational order parameter $t$ for different $\lambda$ values.
 The solid lines are polynomial fits to the data.
 The lower panels (a2)--(e2) show the 
 density-dependence of the orientational order parameter $Q_{6}$.
 In each panel, the different curves correspond to isotherms 
 (top to bottom) $T=0.03, 0.04, 0.05, 0.06, 0.07, 0.08, 0.09, 0.10$.
}
\label{tq-d} 
\end{figure}

These effects of density on the pair correlation function
underlie the evolution of $t$ upon compression, shown in Figs.~\ref{tq-d}(a1)-(e1). Consider
for example the $T=0.04$ isotherm when $\lambda=2/7$. It can be seen that $t$ displays a
non-monotonic dependence on density: it increases upon compression at low
densities, $1.0<\rho<1.22$, decreases over the intermediate density range $1.22<\rho<1.76$, and 
increases again at high densities, $\rho>1.76$. The initial increase at
low densities is associated with the growth of $g(\sigma_1)$. The emergence of structure
associated with the inner (hard) core causes $t$ to decrease at intermediate
densities because the initial, modest growth of $g$ at $r\sim\sigma_0$ causes $|g-1|$ to
decrease with respect to its low-density value of 1 (see Eq.~\ref{top}). Upon further
compression, the growth of $g(\sigma_0)$ above 1 eventually contributes additional area to
the integral of $|g-1|$, causing $t$ to increase. This qualitative behavior of $t$ is
similar for $\lambda=0, 2/7$ and 4/7, and is more pronounced at low $T$. For $\lambda=6/7$,
close to the hard sphere limit, the pronounced growth of the inner peak upon
compression gives rise to a monotonic density dependence of $t$. The case $\lambda=5/7$ is
clearly transitional, with non-monotonic behavior at low temperature changing to
hard-sphere-like monotonic growth of t upon compression at high temperature.
 
Orientational order, as measured by $Q_6$, shows a pronounced dependence on $\lambda$, 
illustrated in Figs.~\ref{tq-d}(a2)--(e2).  When $\lambda=0$ (no hard core), 
$Q_6$ increases monotonically with density for all $T$.
When $\lambda=2/7$, $Q_6$ begins to exhibit non-monotonic behavior upon compression. For 
this particular value of $\lambda$ the trend is very mild, and is best described as a virtual
insensitivity of $Q_6$ to compression, except for an initial increase at low enough densities.
For $\lambda=4/7$ and 5/7, orientational order exhibits a marked non-monotonic dependence on
density, especially at low temperatures. When coupled with the corresponding behavior of $t$,
this corresponds to a water-like structural anomaly, whereby both order metrics decrease
upon isothermal compression. When $\lambda=6/7$, which is close to the hard sphere value
($\lambda=1$), orientational order increases monotonically upon compression. Thus, there exists a 
narrow interval of $\lambda$ within which the ramp fluid shows water-like structural order, 
whereas in the pure ramp ($\lambda=0$) and quasi-hard-sphere limits ($\lambda\sim1$) 
$Q_6$ behaves conventionally upon compression. 

\begin{figure}
\includegraphics[width=7cm,height=18cm]{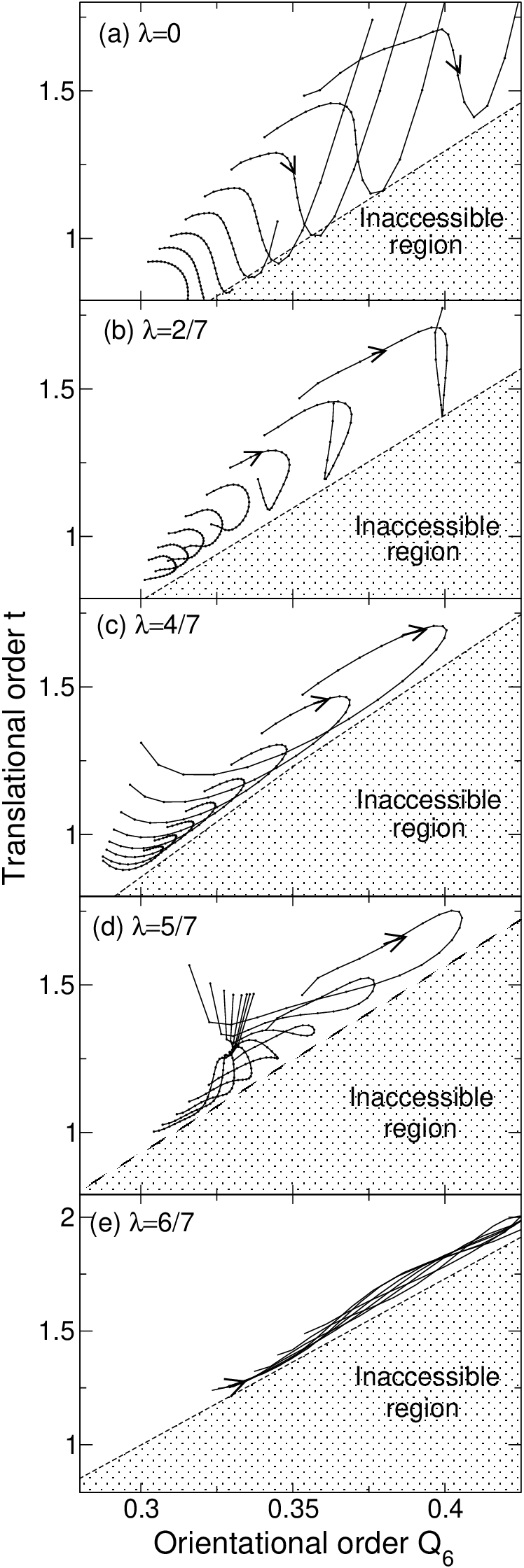}
\caption{Order map of the ramp fluid for different $\lambda$.
  For each $\lambda$, the eight isotherms (right to left) correspond to 
  $T=0.03, 0.04, 0.05, 0.06, 0.07, 0.08, 0.09, 0.10$, 
  and the arrow indicates the direction of increasing density.
}
\label{ordermap}
\end{figure}

Cross-plotting the order metrics against each other generates the order map, whose
evolution as a function of $\lambda$ is depicted in Fig.~\ref{ordermap}. 
For all values of $\lambda$ except
6/7, state points fall on a two-dimensional region, signifying that $t$ and $Q_6$ can be 
varied independently. As is the case for silica and water~\cite{jeffrey01,shell02}, 
we find, for all values of $\lambda$, 
an inaccessible region where no liquid state points can be 
found. In the pure ramp ($\lambda=0$) case, 
the pronounced non-monotonic dependence of $t$ on density gives
rise to isotherms with well-characterized $t$-minima, the locus of which defines the boundary
between the accessible and inaccessible regions of the order map. For $\lambda$ = 2/7, the 
barely discernible non-monotonic dependence of $Q_6$ on density gives rise to loops along
isotherms. The non-monotonic behavior of $Q_6$ is fully developed for $\lambda$ = 4/7. This gives 
rise to an order map with states corresponding to structural anomalies lying on a narrow stripe
of the order map adjacent to the boundary between the accessible and inaccessible regions.
This behavior is strikingly analogous to that of water. The insensitivity of structural
order to temperature, a distinguishing feature of hard spheres, can be clearly seen in 
Fig.~\ref{ordermap}(e) by the virtual collapse of all isotherms in the  $\lambda$ = 6/7 case. 
The transition from water-like to hard sphere order map occurs in the narrow interval 
$4/7<\lambda<6/7$. In particular, for $\lambda$ = 5/7, there is a clear evolution from water-like 
low-$T$ behavior ($T<0.07$) to hard-sphere-like high-$T$ behavior ($T>0.07$). 

\subsection{Thermodynamic, dynamic, and structural anomalies}

We now discuss the regions of the phase
diagram where structural, dynamic and thermodynamic anomalies occur. In water~\cite{jeffrey01},
structural, dynamic, and thermodynamic anomalies occur as nested domes in the ($T, \rho$) or 
($T, P$) planes. Structural anomalies define the outer dome, within which isothermal compression
results in a decrease of both translational and orientational order. Dynamic anomalies
define an intermediate dome, lying entirely within the structural anomalies dome, and
within which isothermal compression leads to an increase in the diffusion coefficient.
Thermodynamic anomalies define the innermost dome, within which water expands when cooled
isobarically. In silica~\cite{shell02}, dynamic anomalies define the outer dome, structural 
anomalies the intermediate dome, and thermodynamic anomalies define the inner dome. Thus, in both
cases negative thermal expansion implies also diffusive and structural anomalies, but in
silica diffusive anomalies occur over a broader range of densities and temperatures than
structural anomalies, the opposite being true in water.

\begin{figure}
\includegraphics[width=10cm,height=8cm]{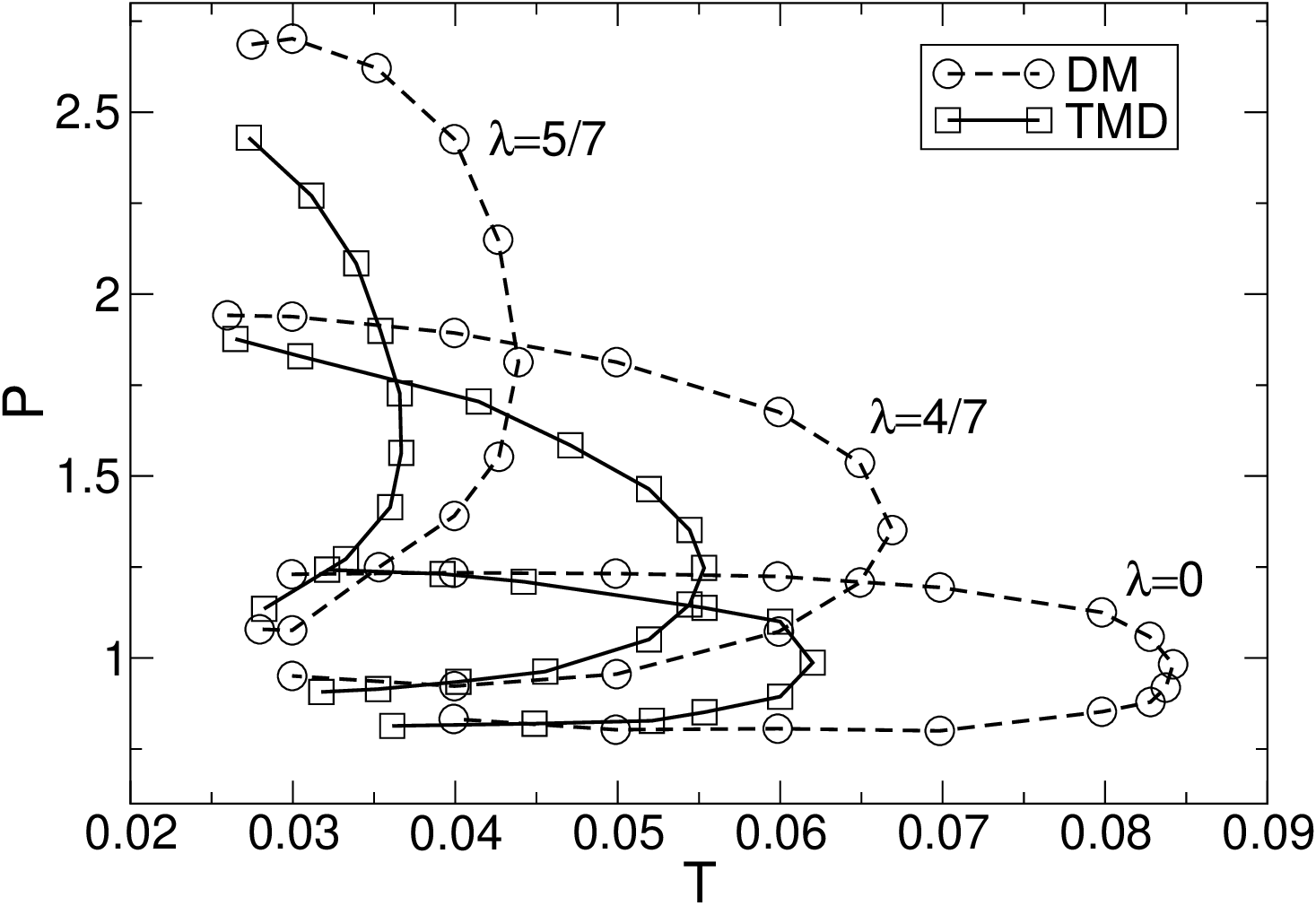}
\caption{
     Loci of thermodynamic and dynamic anomalies
     for ramp fluids with different $\lambda$ values. 
     The region of diffusion anomalies is defined by the loci of 
     diffusion minima and maxima (DM) inside which the diffusivity increases 
     upon isothermal compression.
     The thermodynamically anomalous region is defined by 
     locus of temperatures of maximum density (TMD), 
     inside of which the density increases when the system is heated at 
     constant pressure. 
} 
\label{ptphase} 
\end{figure}

Fig.~\ref{ptphase} shows the loci of dynamic and thermodynamic anomalies for three values of $\lambda$.
Similar to water and silica, in ramp fluids thermodynamic anomalies occur over a narrower
temperature and density range than dynamic anomalies. In other words, if a ramp fluid is at
a state point where it expands when cooled isobarically, its diffusion coefficient
necessarily increases upon isothermal compression. It can be seen that upon increasing $\lambda$,
the range of temperatures where anomalies occurs shrinks, whereas the upper limit of
density (or pressure)  where anomalies can occur increases. The shrinking of the temperature range 
where anomalies occur follows from the fact that increasing 
$\lambda$ makes the fluid progressively hard-sphere-like, and there are no 
anomalies in a hard sphere fluid.

\begin{figure}
\includegraphics[width=16cm, height=6cm]{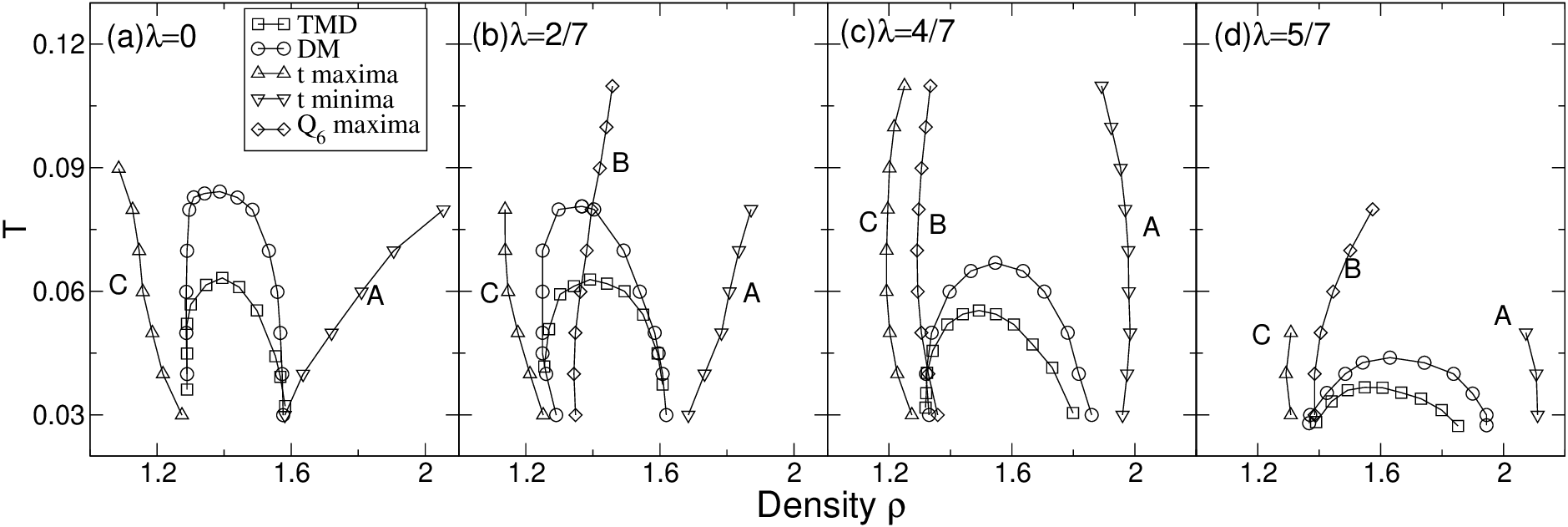}
\caption{
Relationship between structural order and the density and
 diffusion anomalies in the $\rho-T$ plane.
For $\lambda=0$ and 2/7, the domes of dynamic and thermodynamic anomalies
are bounded by loci of $t$ maxima (C) and minima (A), between which isothermal
compression cause a {\it decrease} in translational order. For $\lambda=4/7$ and 5/7, the domes of
dynamic and thermodynamic anomalies are bounded by loci of $Q_6$ maxima (B)
and t minima (A), between which isothermal compression cause a $decrease$ in both translational
and orientational order (structural anomaly). This cascade of anomalies is 
characteristic of water.
}
\label{qttmddm}
\end{figure}

Fig.~\ref{qttmddm} shows the relationship between the loci of dynamic, thermodynamic and structural
anomalies. In water, the low-density and high-density branches of the dome of structural
anomalies correspond to tetrahedrality maxima and translational order minima, respectively.
For the pure ramp case ($\lambda$ = 0), the orientational order increases monotonically with 
density over the range of temperatures explored here. Accordingly, as seen in Fig.~\ref{qttmddm}(a), the
dynamic and thermodynamic anomalies domes are bounded by loci of translational order
extrema (maxima:  line C; minima: line A). Between lines C and A, compression leads to a
decrease in translational order.
For $\lambda$ = 4/7 and 5/7, the locus of orientational 
order maxima (B) provides a low-density bound to the existence of thermodynamic and dynamic
anomalies. Thus, for these two values of $\lambda$, ramp fluids exhibit a water-like cascade of
anomalies (structural, dynamic, thermodynamic).
For $\lambda=2/7$, $Q_6$ maxima are barely discernible and the two order metrics are only weakly
coupled to each other. Accordingly, the locus of weak orientational order maxima (B) is not a 
relevant indicator of dynamic or thermodynamic anomalies.

\section{CONCLUSION} 

In this work we have investigated thermodynamic, dynamic and structural
anomalies in ramp potential fluids, as a function of the ratio $\lambda$ of length scales
corresponding to the inner hard core, and to the outer edge of the ramp. We find that
thermodynamic and dynamic anomalies exist for $\lambda$ = 0, 2/7, 4/7 and 5/7, but not for = 
6/7. As in water and silica, the loci of anomalies form nested domes in the ($T, \rho$) plane, inside 
which the thermal expansion coefficient is negative (inner dome) and the diffusivity
increases upon compression (outer dome). The limit $\lambda$ = 1 corresponds to hard spheres, 
and the absence of anomalies for $\lambda$= 6/7 indicates approach to hard sphere behavior. The 
order map of this family of ramp fluids is water-like at $\lambda$ = 4/7 and hard-sphere-like at $\lambda$ = 6/7. 
Thus, by varying the ratio of characteristic length scales, the family of ramp potentials spans the
range of liquid behavior from hard spheres to water-like.

These findings show that orientational interactions are not necessary for the existence of
thermodynamic, dynamic, or structural anomalies. Instead, water-like behavior apparently
emerges in this spherically-symmetric family of fluids through the existence of two
competing length scales. Although thermodynamic and dynamic anomalies exist almost over the
entire range of the control parameter, the combination of thermodynamic and dynamic
anomalies plus a water-like order map occurs over a narrow range of $\lambda$. It is interesting to
note that a distinguishing feature of water is the fact that the ratio of radial distances
to the first and second peaks of the oxygen-oxygen pair correlation function is not $\sim1/2$,
as in simple liquids, but $\sim0.6$. This is close to 0.571 ($\lambda$ = 4/7, the ratio of 
$\sigma_0$ to $\sigma_1$
that gives rise to water-like structural, dynamic and thermodynamic anomalies). 
In water, isothermal compression
pushes molecules from the second shell towards the first shell, gradually
filling the interstitial space~\cite{scio92}. Likewise, in the ramp potential, 
isothermal compression pushes molecules from the soft core ($\sigma_1$) to the 
hard core ($\sigma_0$).
Further work is
needed to establish whether a ratio of competing length scales close to 0.6 is generally
associated with water-like anomalies in other core-softened potentials, for example linear
combinations of Gaussian~\cite{still76} potentials. In this work we used the terminology water-like to
denote structural, diffusion, and density anomalies. The increase in water's isothermal
compressibility upon isobaric cooling, another of this liquid's canonical anomalies, is also
trivially captured by the ramp potential, because thermodynamic consistency arguments~\cite{sastry96}
mandate that the compressibility increase upon cooling whenever there exists a
negatively-sloped locus of density maxima in the ($P, T$) plane.

The ramp potential, when supplemented by explicit~\cite{Jagla01,xu05} or mean-field 
attractions~\cite{Jagla99}, 
gives rise to liquid-liquid immiscibility and a critical point distinct from the one associated
with the vapor-liquid transition. A liquid-liquid transition has been observed
experimentally in phosphorus~\cite{katay00,katay04}, n-butanol~\cite{kurita05} and triphenyl 
phosphite~\cite{kurita04}, 
and strong experimental evidence consistent with liquid-liquid immiscibility also exists for water
\cite{mishima98,mishima00,mishima02}.
Computer simulations of silicon~\cite{sri}, silica~\cite{saika,lacks00}, carbon~\cite{glosli99} 
and water~\cite{poole,harri97,yamada03,poole05,yamada02,brovch03,brovch05} 
also indicate the presence of a liquid-liquid transition. A systematic study of
the effects of $\lambda$ and the ratio of characteristic energies ($U_1$ and the attractive well 
depth) on the existence of a liquid-liquid transition, the positive or negative slope of the line
of first-order liquid-liquid transitions in the ($P, T$)  plane, and the relationship, if any
\cite{franzese}, between the liquid-liquid transition and density anomalies, would shed important 
new light on the phenomenon of liquid polyamorphism~\cite{angellPCCP,poole97,yarger04}.

It is generally accepted that strong orientation-dependent interactions underlie many of
the distinctive properties of associating, network-forming liquids such as water. Atomic
liquids, on the other hand, exhibit simpler behavior, and in particular do not show
structural, thermodynamic, or dynamic anomalies of the type discussed here. In this work we
have shown that key properties of these apparently distinct categories of liquids can be
bridged systematically by varying the ratio of two length scales in a family of
spherically-symmetric potentials in which orientation-dependent interactions are absent by
design. What other spherically-symmetric potentials, in addition to those possessing
competing length scales, may give rise to water-like anomalies, is among the interesting
questions arising from this study that we will pursue in future work.

\section{Acknowledgments}
We thank NSF grants CHE~0096892 and CHE~0404699 for support and Yeshiva University
for CPU time.

\newpage

\newpage

\newpage

\newpage

\newpage

\newpage

\end{document}